\documentstyle[aps,preprint,prl,epsf]{revtex}
\tightenlines

\newcommand{\s}{\mbox{$\sigma$}}
\newcommand{\be}{\begin{equation}}
\newcommand{\ee}{\end{equation}}
\newcommand{\ua}{\mbox{$\uparrow$}}
\newcommand{\da}{\mbox{$\downarrow$}}
\newcommand{\k}{{\bf k}}
\newcommand{\e}{\mbox{$\epsilon$}}

\begin{document}

\title{\Large{\bf Coexistence of spin-triplet superconductivity and
ferromagnetism induced by the Hund's rule exchange
}}

\vskip0.5cm 

\author{ J. Spa\l{}ek,  P. Wr\'obel }

\address{Marian Smoluchowski Institute of Physics, Jagiellonian University, ulica Reymonta 4, \\
30-059 Krak\'{o}w, Poland\\}

\author{  and W. W\'ojcik }

\address{Institute of Physics, Tadeusz Ko\'sciuszko Technical University,\\
ulica Podchor\c{a}\.zych 1, 30-084 Krak\'ow, Poland }

\maketitle

\vskip0.5cm


\begin{abstract}

\noindent We discuss general implications of the local spin-triplet pairing
among correlated fermions that is induced by the Hund's rule coupling in
orbitally dege\-ne\-rate systems. 
The quasiparticle energies, the magnetic moment,  
and the superconducting gap are determined for  
principal superconducting phases, in the situation with the exchange field
induced by both the local Coulomb and the Hund's rule exchange interactions. 
The phase diagram, as well as the evolution in an applied magnetic field of the
spin-triplet paired states near the Stoner threshold is provided for a model
two-band system. 
The appearance of the spin-polarized superconducting phase makes the Stoner
threshold a hidden critical point, since the pairing creates a small but
detectable uniform magnetization.  
The stability of the superconducting state against the ferromagnetism with an
alternant orbital ordering appearing in the strong-coupling limit is also
discussed. 

\end{abstract}

\vskip0.5cm



\newpage

\noindent {\Large {\bf 1. Introduction
} }

\vskip0.5cm 

The discovery of superconductivity in $Sr_2RuO_4$ \cite{1}, and particularly of
its coexistence with ferromagnetism in $UGe_2$ \cite{2}, $ZrZn_2$ \cite{3}, and
$URhGe$ \cite{4} showed clearly that the long awaited spin-triplet superconducting
state is realized in Nature. 
The above three  systems have a weak and an itinerant nature of the
electrons involved in both ferromagnetism and the pairing. 
Therefore,  the models of correlated electrons generalized to orbitally
degenerate systems (such as the degenerate Hubbard model) should be a starting
point in theoretical considerations of the phases involved, since they are
certainly applicable to description of the itinerant magnetism. 
Furthermore, they should also be regarded also as a providing 
proper pairing mechanism for those systems,  
since the superconductivity disappears at  high applied pressure together
with ferromagnetism and hence, it is unlikely that it is caused by a
nonmagnetic mechanism (e.g. by the electron-phonon coupling), which should not
be influenced by the presence (or absence) of ferromagnetism to such an 
extent. 
Moreover, the superconductivity appears together with/or 
inside the ferromagnetic phase only
when magnetism is rather weak (magnetic moment is small), i.e. when the system
is susceptible to a local exchange-field enhancement (by  formation of a
pair bound states in spin-triplet state). 
In other words, the spin-triplet pairing should be enhanced in the vicinity of
the 
Stoner critical point provided that the quantum spin 
fluctuations do not introduce too strong
scattering of the individual carriers. 
In the present article we extent our original treatment \cite{5,6} of
spin-triplet superconductivity and discuss its coexistence with the itinerant
ferromagnetism within a single mechanism responsible for the appearance  
of both of them - the Hund's rule ferromagnetic exchange among correlated and
orbitally degenerate d states of narrow-band electrons. 
The structure of the paper is as follows. 
In the next two section we build a formal structure of the theoretical
approach. 
Namely, we introduce the concept of the real-space spin-triplet pairing, as
well as generalize the Nambu-Bogolyubov-deGennes formalism to the situation
with spin-triplet pairing. 
In Section 4 we discuss the spin-triplet pairing below the Stoner threshold. 
The most important message there is that the Stoner critical point is actually
a hidden critical point in the spin-triplet paired state. 
In Section 5 we present analytic estimates of the superconducting gaps and 
critical temperatures in the weakly ferromagnetic  states. 
Finally, in Section 6 we provide the phase diagram in the strong correlation 
limit and in particular, discuss the orbital ordering as well.

\vskip1.0cm 

\noindent {\Large {\bf 2. Real space pairing induced by the local ferromagnetic
exchange
}} 

\vskip0.5cm 

We start from an extended Hubbard model of correlated and orbitally degenerate
narrow-band electrons represented by the parametrized Hamiltonian
$$
{\cal{H}} = \sum_{ijll'\s}{''} t_{ij}^{ll'} a_{i  l \s}^{\dagger}a_{j  l' \s} 
+U\sum_{il}n_{il\ua}n_{ il\da} + \frac{1}{2} U' \sum_{ill'\s \s '}{'}
n_{il\s}n_{il'\s '} $$
\be 
-J \sum_{ill'}{'} \left( {\bf S}_{il} \cdot {\bf S}_{il'} + \frac{3}{4}
n_{il}n_{il'} \right) + J \sum_{ill'}{'} a_{il\ua}^{\dagger} a_{il\da}^{\dagger}
a_{il'\da} a_{il'\ua} . 
\label{1}
\ee 
In this Hamiltonian the first term describes the electron hopping between the
atomic sites $i$ and $j$ and between the orbitals $l$ and $l'$; the double
primed summation means that both $l \neq l'$ and $i \neq j$. 
The next two terms describe the direct Coulomb interactions, intra- and
inter-orbital terms, respectively. 
The last two terms represent the Hund's rule ferromagnetic exchange and the
pair hopping, respectively. 
In what follows we are interested in the spin-triplet correlations and pairing,
so the first task is to construct an effective Hamiltonian with pairing
renormalized by the Coulomb interactions $U$ (the third and the last terms play
only a minor role, at least in the weak-coupling regime). 
This procedure has been carried out earlier \cite{5} for the two-band case
within the auxiliary (slave) boson formalism in the saddle-point 
approximation. 
However, to introduce the starting effective Hamiltonian in an explicit form in
a weak-coupling limit we 
introduce real-space spin-triplet pairing operators via the following 
relations 
\be 
\left\{ 
\begin{array}{c}
A_{i1ll'}^{\dagger} = a_{il\uparrow}^{\dagger} 
a_{il'\uparrow}^{\dagger} \; \; \; \; \; \; \; \; \; \; \; \; \; \; \; \; \; for \; \; \; S_l^z+S_{l'}^z \equiv m=1, \\  
A_{i0ll'}^{\dagger} = \frac{1}{\sqrt{2}} \left( 
a_{il\uparrow}^{\dagger} a_{il'\downarrow}^{\dagger}  + 
a_{il\downarrow}^{\dagger} a_{il'\uparrow}^{\dagger} \right) \; \; \; \; \; \;
\; \; 
for \; \; \; m=0, \\  
A_{i-1ll'}^{\dagger} = a_{il\downarrow}^{\dagger}
a_{il'\downarrow}^{\dagger} \; \; \; \; \; \; \; \; \; \; \; \; \; \; \; \;  \; \; \; \; \; \; \; \; \; \; \; \; for \; \; \; m=-1, 
\end{array} 
\right. 
\label{2}
\ee 
and neglect as irrelevant the third and the fifth term in (\ref{1}). 
In effect, we obtain 
\be 
{\cal{H}} = \sum_{ill' \s}{'} t_{ij}^{ll'} a_{il\s}^{\dagger} a_{il'\s} + U
\sum_{il} n_{il\ua}n_{il\da} - J \sum_{imll'} A_{imll'}^{\dagger} A_{imll'} . 
\label{3} 
\ee 
The first term provides the hybridized band states, the second the repulsion
between electrons on the same orbital but with opposite spins (the Hubbard
term), and the third 
introduces local spin-triplet correlations for electrons located on the
orbitals $l$ and $l'$, $l \neq l'$. 

The simplest solution of the model is to make the Hartree-Fock approximation. 
We will study the solution for which the ferromagnetic moment $<S_l^z> =
<n_{il\ua} - n_{il\da}>/2$, and the anomalous superconducting averages
$\Delta_{ll'm} \equiv 2J(d-1) <A_{imll'}^{\dagger}>$ are nonzero and represent
a stable solution. 
We assume additionally that the bands are equivalent, i.e. put $<S_l^z> =
{\bar{S^z}}$ and $\Delta_{ll'm} = \Delta_m$. 
Such a procedure leads to the Hartree-Fock Hamiltonian of the form 
$$ 
{\cal{H}} = \sum_{ijll'\s}{''}  t_{ij}^{ll'} a_{il\s}^{\dagger} a_{jl'\s} -
4J(d-1) {\bar{S^z}} \sum_{il} S_{il}^z + Jd(d-1) N{\bar{S^z}}^2 $$ 
\be 
- \sum_{im l\neq l'} \left( \Delta_m A_{im ll'}^{\dagger} + \Delta_m^*
A_{imll'} \right) + Jd(d-1) |\Delta_m |^2 N - 2U{\bar{S^z}} \sum_{il} S_{il}^z
+ UdN{\bar{S^z}}^2, 
\label{4}
\ee 
where $d$ is the orbital degeneracy and $N$ is the number of atomic sites. 
Equivalently, we can write 
$$ 
{\cal{H}} = \sum_{ijll'\s}{''}  t_{ij}^{ll'} a_{il\s}^{\dagger} a_{jl'\s} -
2[U+2J(d-1)] {\bar{S^z}} \sum_{il} S_{il}^z  $$ 
\be
- \sum_{im l\neq l'} \left( \Delta_m A_{im ll'}^{\dagger} + \Delta_m^*
A_{imll'} \right) +  
+ \{ [U+J(d-1)]{\bar{S^z}}^2 + \frac{ |\Delta_m |^2}{2J(d-1)} \} Nd.
\label{5}
\ee 
We see that the quantity $I = 2[U + 2J(d-1)]$ is the magnetic coupling constant
and the coupling constant for spin-triplet pairing is $J(d-1)$. 
We have a ferromagnetism coexisting with spin-triplet paired phase if both
${\bar{S^z}}$ and at least one of the gap parameters $\Delta_m$ $(m=+1,0,-1)$
are nonzero simultaneously for the energetically stable solution. 
In what follows we provide the solution of the Hamiltonian (\ref{5}) in a model
two-band situation, i.e. neglect the hybridization of the bands (put $t^{12}
= 0$). 
As long as the d-fold degenerate bands regarded as almost equivalent, such
two-band model should catch the essential qualitative features of the
solutions.

\vskip1.0cm 

\noindent {\Large {\bf 3. Spin-triplet superconducting state 
}} 

\vskip0.5cm 

In the absence of spin-triplet superconductivity a  two-band system is 
paramagnetic below the Stoner threshold, i.e. when $\rho ( \e_F )I <1$, where
$\rho (\e_F )$ is the density of states at the Fermi energy $\e_F$. 
If we have a degenerate two-band system with flat density of states and of the
bandwidth $W$ each, then the condition takes the form $2I/W <1$. 

To solve the system of self-consistent equations for $\Delta_m$, ${\bar{S^z}}$,
and the chemical potential $\mu$ we have  generalized \cite{5,6,7} 
the Nambu-Bogolyubov-de Gennes 
notation and have constructed $4 \times 4$ matrix representation of the
Hamiltonian by defining the composite creation operators of the form 
 ${\bf f}_{\k}^{\dagger} = ( f_{\k 1 \ua}^{\dagger}, f_{\k 1 \da}^{\dagger}, 
f_{-\k 2 \ua}, f_{-\k 2 \da} )$, and the annihilation operators
as ${\bf f}_{\k} = ({\bf f}_{\k}^{\dagger})^{\dagger}$. 
We have 
\be 
{\cal{H}} = \sum_{\k} {\bf f}_{\k}^{\dagger} {\bf A} {\bf f}_{\k} + 
\sum_{\k} E_{\k 2} +  I({\bar{S^z}})^2N + \frac{|\Delta_m|^2}{J(d-1)}N , 
\label{6}
\ee 
with 
\be 
{\bf A} = \left( 
\begin{array}{cccc} 
E_{\k 1} - I{\bar{S^z}} , & 0, & \Delta_1 , & \Delta_0  \\ 
0 , & E_{\k 1} + I{\bar{S^z}} , & \Delta_0 , & \Delta_{-1}  \\ 
\Delta_1 , & \Delta_0 , & -E_{\k 2} + I{\bar{S^z}} , & 0 \\ 
\Delta_0 , & \Delta_{-1} , & 0 , & -E_{\k 2} - I{\bar{S^z}} 
\end{array} 
\right) . 
\label{7}
\ee 
The quantities $E_{\k 1} \equiv E_{\k 1} - \mu$ and $E_{\k 2} \equiv E_{\k 2} -
\mu$ are the band energies (with the chemical potential as a reference point). 
This matrix  can be brought to a diagonal form analytically for the
interesting us here case with $\Delta_0=0$ (the phase with $\Delta_0 = 0$ is
almost always energetically unstable). 
In such situation we obtain the following four eigenvalues 
\be 
\lambda_{\k \s 1,2} = \frac{1}{2} \left( E_{\k 1} - E_{\k 2}\right) \mp \left[
\frac{1}{4} \left( E_{\k 1} + E_{\k 2} - \s I{\bar{S^z}} \right)^2 +
|\Delta_{\s}|^2 \right]^{1/2}, 
\label{8}
\ee 
where the sign $(\mp )$ corresponds to the label $(1,2)$ of the eigenvalues
$\lambda_{\k \s 1,2}$. 
The quasiparticle 
spectrum separates into a pair of spin subbands with the 
 spin splitting $\delta \equiv \lambda_{\k
\da i} - \lambda_{\k \ua i}$, determined mainly by the exchange field, 
since $I$ is substantially larger than $J$. 
The spectrum is fully gapped if both $\Delta_{1} \equiv 
\Delta_{\ua \ua}$ and $\Delta_{-1} \equiv \Delta_{\da \da}$ are
nonzero; this phase is called in analogy to superfluid helium as the
anisotropic A phase (in general, $\Delta_{\ua \ua} \neq \Delta_{\da \da}$ 
in the ferromagnetic phase). 
However, if only one component $(\Delta_{\ua \ua})$ 
of the gap is nonzero,  then 
$ \lambda_{\k \da 1} = - E_{\k 2} + I{\bar{S^z}}$ and 
 $ \lambda_{\k \da 2} =  E_{\k 1} + I{\bar{S^z}}$. 
This means that the minority spin spectrum is ungapped and will thus produce a
nonzero linear specific heat $\gamma_{\da}T$, with $\gamma_{\da} \sim
\rho_{\da} (\e_F )$ if the bands are symmetric with respect to their middle
point (i.e. respect the electron-hole symmetry). 
The phase with $\Delta_{\ua \ua} \neq 0$, $\Delta_{\da \da} =0$ will be 
called the A1 
phase (we take here a convention that in ferromagnetic phase magnetic moment
${\bar{S^z}} > 0$ and $\Delta_{\ua \ua} \neq 0$; a physically equivalent 
but distinct state is 
that with $(-{\bar{S^z}})$ and $\Delta_{\da \da} \neq 0$). 
Note also that the appearance of the A1 phase does not necessarily mean that we
are in the ferromagnetically saturated phase, i.e. 
with $<S^z> = n/4$, where $n$ is 
the band filling, defining here as the number of electrons per site. 

One should also mention that in the applied magnetic field $B \neq 0$, all the
above results are valid except we have to make a replacement $I{\bar{S^z}}
\rightarrow I{\bar{S^z}}+ \mu_B B$, where $\mu_B$ is the Bohr magneton. 

The Bogolyubov quasiparticle operators can also be  calculated \cite{6} when
diagonalizing (\ref{6}). 
In this paper we provide the explicit results only for the case $E_{\k 1} =
E_{\k 2} = E_{\k}$, since they have a simple interpretation then. 
Namely, under this condition the quasiparticle energies take 
the usual form 
\be 
\lambda_{\k \s 1,2} = \pm \left( E_{\k \s}^2 + |\Delta_{\s}|^2 \right)^{1/2}
\equiv \pm \lambda_{\k \s} , 
\label{9} 
\ee 
with $E_{\k \s} = E_{\k} - \s (\mu_B B + I {\bar{S^z}})$. 
Also the quasiparticle operators corresponding to the eigenvalues $\pm
\lambda_{\k}$ have respective forms "$+$" for $\alpha$ quasiparticles, "-" for
$\beta$: 
\be 
\left( 
\begin{array}{c} 
\alpha_{\k \s} \\
\beta_{-\k \s}^{\dagger}
\end{array} 
\right) = \frac{1}{\sqrt{2}} 
\left( 
\begin{array}{cc} 
u_{\k}^{(\s )}, &  v_{\k}^{(\s )} \\ 
- v_{\k}^{(\s )}, &  u_{\k}^{(\s )} 
\end{array} 
\right) 
\left(
\begin{array}{c} 
f_{\k 1\s} + f_{-\k 2\s}^{\dagger}  \\
f_{\k 1\s} - f_{-\k 2\s}^{\dagger}  
\end{array}
\right) ,
\label{10}
\ee 
with the coherence factors 
\be 
\left( 
\begin{array}{c} 
u_{\k}^{(\s )} \\ 
v_{\k}^{(\s )} 
\end{array} 
\right) = 
\frac{1}{\sqrt{2}} 
\left( 
\begin{array}{c} 
1 + \frac{\Delta_{\s}}{\lambda_{\k \s}} \\ 
1 - \frac{\Delta_{\s}}{\lambda_{\k \s}}
\end{array} 
\right) . 
\label{11} 
\ee 
We see that the equations for the quasiparticles in the spin subband $\s$ have
in this limit the same form as in the BCS case. 
In other words, we have two gaps in an anisotropic A phase 
in the system and they are induced by the
presence of the molecular field (when $B=0$). 
Hence in the paramagnetic state $({\bar{S^z}} =0)$ we should have an isotropic
A phase $( \Delta_{\ua \ua} = \Delta_{\da \da} = \Delta \neq 0, \; 
\Delta_{0} = 0)$ as the stable phase. 
We shall see that this is {\it not} always the case, i.e. the superconducting
pairing may produce a nonzero  spin polarization even below the
Stoner threshold, as we discuss next.

\vskip1.0cm 

\noindent {\Large {\bf 4. Spin-triplet paired state below the Stoner threshold:
Phase diagram and a hidden critical point
}} 

\vskip0.5cm

One should note that for $d$-electron systems $J$ is of the order $0.1-0.3U$. 
In the numerical calculations we therefore take $J = 0.25I$. 
If we take also the density of states of the single band as $(1/W)$ we have
that if the Stoner threshold for the onset of ferromagnetism is reached when 
 $J/W = 0.125$. 
In Fig.1 we have plotted the ground state energy (in units of $W$) for the A,
A1, and normal $(\Delta_m \equiv 0)$ states as as a function of applied
magnetic field B (all the energies and the parameters are expressed in 
units of $W$). 
The energy difference between A and A1 phases is small and at the applied field
of the order of $\mu_B B = 5.10^{-4} W$ A $\rightarrow$ A1 transition 
takes place  
(for $W = 1eV$  this critical field is $\sim 50T$). 
For this applied field magnitude the gap anisotropy is $\Delta_{\ua \ua}/
\Delta_{\da \da} \sim 3$, as displayed in lower panel in Fig.2 (not that the
$\Delta_{\ua \ua} (B)$ dependence is almost the same in both A and A1  
states).  

In Fig.2 we display also the value of the magnetic moment per orbital
$(<S_l^z>)$ and (in the inset) the field dependence of the chemical potential
in both A and A1 paired states. 
Again, the magnetic moment in the A1 state (dashed line) follows essentially
the same straight-line dependence ${\bar{S^z}}(B)$ for the both paired states. 
In this sense, magnetic properties are not influenced much by the pairing. 
In view of this last feature of the solution it is not strange that the A1
phase is stable even though the system is not yet magnetically saturated. 

One very interesting feature of our mean-field approach should be mentioned. 
Namely, the ${\bar{S^z}}(B)$ in the paired state dependence {\it does not}
approach exactly the value ${\bar{S^z}} =0$ for $B=0$, even though the system is
below the Stoner threshold. 
The effect is small to be come visible in the lower left hand corner of the upper
panel, but it is certainly well above the numerical accuracy of the results. 
To test our conjecture that the pairing itself may introduce a uniform 
ferromagnetic 
polarization we have calculated this {\it remament} value of the spin magnetic
moment in the field $B=0$ when approaching the Stoner critical point. 
The result is displayed in Fig.3. 
We observe a beautiful critical dependence of the moment as we approach the
Stoner point. 
So, indeed, the pairing washes out the critical Stoner point, i.e. makes it a
hidden point. 
It is interesting to ask to what extent the quantum critical fluctuations can
change this mean-field result. 
The result also means that the superconducting coherence length becomes
infinite at the Stoner point. 
It remains to be seen whether it is unbound whenever the A1 phase sets in. 

The results displayed in Fig.3 contain also one additional feature exhibited 
in the inset. 
Namely, the inset shows that if no pairing were present then the mean-field 
para-ferro-magnetic transition would be discontinuous (for the assumed 
constant density of states) and directly to the saturated state. 
The pairing smears out this discontinuity and therefore, we have an extended
critical regime for $J/W \rightarrow 0.125$. 
Additionally, because of the absence of the critical point for ${\bar{S^z}}
(J)$ dependence it is difficult to say where the ferromagnetism disappears as a
function of e.g. pressure. 
This is exactly what is actually observed for the newly discovered
superconducting ferromagnets \cite{2,3}.

The fact that the spin-triplet pairing can induce a weak ferromagnetic ordering
must mean that the coherence length $\xi$ of the paired states is larger than
the classical distance $(V/N)^{1/3}$ between the electrons in this system of
volume $V$ containing $N$ electrons. 
The overlap between the Cooper pairs effectively induces a spin-spin
interaction, which can be understood in the following manner.  
The superconducting gap creates effective magnetic field $H_{jm} = \chi_{ji}
\Delta_{mi}$, which in turn induces magnetic moment $M_{i} \sim \chi_{ji}
\Delta_{mi}$ ($\chi_{ji}$ is the superconducting susceptibility) and in turn, 
a negative contribution to magnetic energy $\sim ({\bar{S^z}})^2$. 

Of particular interest is the stability of the paired states when approaching
the Stoner threshold from the paramagnetic side. 
The border line between the A and A1 phases is drawn as a solid line in Fig.4. 
The A phase disappears exactly at the Stoner point, but the A1 survives. 
The reason why only  A1 phase can survive at the critical point is very
simple. Namely, the magnetic susceptibility is infinite at this point, so even
a weak field induces total polarization. 
However, strictly speaking A1 phase should not be the stable state since in the
magnetically saturated state there is no way we can increase the polarization
locally to form Cooper pairs as proper bound states. 
This is the reason to assume that the superconducting coherence length becomes
infinite at the Stoner point.

\vskip1.0cm 

\noindent {\Large {\bf 5. Spin-triplet state in a weak ferromagnetic state: \\ 
Analytic estimates
}} 

\vskip0.5cm

We discuss now the situation for a weak itinerant (Stoner-Wohlfarth)
ferromagnet, i.e. the system for which $\rho (\e_F )I$ is above but close to
unity. 
In the mean-field approximation, the equation for magnetic moment
$m=2{\bar{S^z}}$ is determined from the Landau-type expansion, which is
obtained from the low-temperature expansion \cite{8} 
\be 
m = I\rho (\e_F ) m + \frac{I^3}{24} \left[ \frac{3\rho ' (\e_F)^2}{\rho (\e_F
)} - \rho '' (\e_F ) \right] m^3 = 0 , 
\label{12} 
\ee 
where $\rho '$ and $\rho ''$ are the corresponding derivatives of $\rho (\e )$
taken at $\e = \e_F$. 
The nonzero solution is thus of the form 
\be 
{\bar{S^z}} = \frac{1}{2} \left[ \frac{I \rho ( \e_F ) - 1}{B} \right]^{1/2} , 
\label{13} 
\ee 
with 
\be 
B = I^3 \frac{\rho (\e_F )}{8} \left[ \frac{\rho ' (\e_F )^2}{\rho (\e_F )^2}
- \frac{\rho '' (\e_F )}{\rho ( \e_F )} \right] . 
\label{14} 
\ee 
In a similar fashion, one obtain the following expression for the Curie
temperature 
\be 
T_C = \frac{\sqrt{6}}{\pi} \left[ \frac{\rho ' (\e_F )^2}{\rho (\e_F )^2}
- \frac{\rho '' (\e_F )}{\rho ( \e_F )} \right]^{-1/2} 
\left[ \frac{I \rho ( \e_F ) - 1}{ I \rho ( \e_F ) } \right]^{1/2} . 
\label{15}
\ee 
Obviously, $T_C$ express the critical temperature for ferro- to para-magnetic
phase transition. 
The magnetization diminishes with temperature in the low-temperature regime
according proportionally to $T^2$ as observed in $URhGe$ \cite{4}; this proves
directly that these systems are weak itinerant ferromagnets (the standard
contribution due to the spin wave excitations is $\sim T^{(2n+1)/2}$, with
$n=1,2,...$). 

In order to estimate the value of the superconducting gap in the A phase we use
the corresponding BCS equation, which for the simplest case with 
$E_{\k 1} = E_{\k 2}$ reads 
\be 
1 = \frac{J}{N} \sum_{\k} \frac{1}{2 \lambda_{\k \s}} \tanh \left(
\frac{\lambda_{\k \s}}{2k_B T} \right) . 
\label{16}
\ee 
To estimate gap $\Delta_{\s}$ at $T=0$ and the temperature $T_S$ of the
transition to the superconducting state we assume that the dispersion relation
in the spin subbands $\s $ is linear, i.e. $E_{\k \s} - \mu \simeq v_{\s} k$,
where $v_{\s}$ is the Fermi velocity in that subband. 
Then, making the usual BCS-type approximation we obtain the equation for
$\Delta_{\s}$ at $T=0$: 
\be 
1 = J \rho_{\s} \int_{-k_{m \s}}^{k_{m \s}} \frac{d^3 (v_{\s} \k ) }{\sqrt{(v_{\s}
k)^2+ \Delta_{\s}^2}} , 
\label{17} 
\ee 
where the density of states at the Fermi level is $\rho_{\s} = 12 \e_{F
\s}^2/W$, $\e_{F\s }$ is the Fermi energy for the quasiparticles in the $\s - $
subband. 
We perform the integration only within the spin-split region of the bands,
which extends from $-I {\bar{S^z}}$ to $+I {\bar{S^z}}$. 
Therefore, the border  wave vector is determined from the relation $\pm v_{\s}
k_m = I {\bar{S^z}}$ ($"+"$ for electrons, $"-"$ for holes). 
In result, the zero-temperature gap  $\Delta_{\s}$ takes the form 
\be                                                 
\Delta_{\s} = I {\bar{S^z}} \exp \left( - \frac{1}{\rho_{\s} J} \right) . 
\label{18} 
\ee 
That gap vanishes identically if we reach the Stoner point, at which
${\bar{S^z}} = 0$. 
Likewise, the estimate of $T_S$ is determined from the equation 
\be 
1 = J \int_{-k_{min}}^{k_{min}} d^3k \frac{ \tanh \left( \frac{\hbar v_F k - 2I
{\bar{S^z}}}{2k_BT_S} \right) }{\hbar v_F k - I {\bar{S^z}} } , 
\label{19} 
\ee 
where we have taken $v_{\s} \simeq v_F$ is the Fermi velocity in the paramagnetic
phase and $k_{min}$ is determined from the equation $v_F~k_{min} = I
{\bar{S^z}}$. 
As a result, we obtain 
\be 
T_S \simeq 2.26 I {\bar{S^z}} \exp \left( - \frac{1}{J \rho } \right) , 
\label{20} 
\ee 
where $\rho \simeq ( \rho_{\ua} + \rho_{\da})/2$. 
From these expressions we can estimate the gap ratio 
\be 
\frac{\Delta_{\ua}}{\Delta_{\da}} \sim \exp \left[ \left( \frac{2I}{3J} \right)
\frac{\rho ' (\e_F )}{\rho ( \e_F )^2}  {\bar{S^z}} \right] . 
\label{21} 
\ee 
A  flat density of states favors isotropic A-phase solution for $B=0$
(cf. Sec.4), whereas for the steep density of states near $\e_F$ we observe a strong
anisotropy. 
Additionally, the ratio increases exponentially with the increasing 
magnetic moment. 
Also, formulae (\ref{15}) and (\ref{20}) allow for a determination of the
$T_S/T_C$ ratio explicitly. 
One can notice immediately that this ratio must be small for weak itinerant
magnet, for which  $I\rho \approx 1$, and then $J \rho \sim 0.2$. 
So, the two critical temperatures can differ by two orders of magnitude easily
(the actual ratio in the newly discovered superconducting ferromagnets
\cite{2,4} is above 
30 in the applied-pressures regime placing the systems not too close to the 
Stoner threshold). 

To summarize, this Section we have three energy scales in the system: (i) the 
$eV$ energy 
scale of $U$ and $J$, (ii) the scale of the exchange splitting
$(2I{\bar{S^z}})$  and associated with it the value of $T_C$, and (iii) the
magnitude of the superconducting gaps $(\Delta_{\ua \ua}, \Delta_{\da \da} )$ 
and associated with it the value of $T_S$. 
Both the weak itinerant-electron ferromagnetism and the spin-triplet 
superconductivity 
are induced by a single mechanism - the local ferromagnetic exchange
interaction. 
The pairing is induced by the exchange interaction itself. 
In this respect our mechanism belongs to the same class of models as the $t-J$
model \cite{9}, for which, the spin-singlet pairing in that case, 
is induced by the kinetic exchange (superexchange). 
The appearance of the exchange-induced paired state is here as natural, as the
presence of ferromagnetism above the Stoner threshold; the phase of
superconducting ferromagnet  is more favorable energetically than either the
purely ferromagnetic or the spin-triplet superconducting states. 
The effect of exchange field is  stronger than that coming from the spin
fluctuations \cite{10}.  
In a way, our approach represent a simplest approach to the spin-triplet
superconductivity, in a complete parity with the Stoner theory of
ferromagnetism.

\vskip1.0cm 

\noindent {\Large {\bf 6. Spin-triplet pairing for strongly correlated
electrons: Role of ferromagnetic superexchange and orbital \\ 
ordering 
}} 

\vskip0.5cm 

The approach in the preceding Sections is based on the notion that electrons in
a narrow band are weakly correlated (i.e. placed physically  
close to the Stoner boundary for ferromagnetism). 
The question arises what would happen if the particles are strongly 
correlated? 
This question is important because the answer in the affirmative would provide
a strong indication that the mechanism is quite universal and hence, is
relevant in the most interesting (and difficult) regime of intermediate
correlations $(U \sim W)$. 

The simplest situation arises for the quarter-filled doubly degenerate band,
for which we have a ferromagnetuc insulator with orbital ordering \cite{7}. 
We have applied the same type of formalism to the case with filling around the
quarter filling and have obtained the following Hamiltonian, with the help of
which one can study the spin-triplet pairing 
\be 
{\cal{H}} = \sum_{ijl \s}{'} t_{ij} b_{il\s}^{\dagger}b_{jl\s} - \frac{2}{K-J}
\sum_{ijk} \sum_{m=-1}^{1} t_{ij}t_{jk} B_{ijm}^{\dagger}B_{jkm} , 
\label{22} 
\ee 
where $b_{il\s}^{\dagger}$ and $b_{jl\s}$ represent the so-called projected
creation and annihilation operators, e.g. 
\be 
b_{i1\s}^{\dagger} = a_{i1\s}^{\dagger} (1 - n_{i1{\bar{\s}}})(1 - n_{i2\s}) 
(1 - n_{i2{\bar{\s}}}), \; \; \; etc. 
\label{23}
\ee 
The pairing operators are 
\be 
\left\{ 
\begin{array}{c}
B_{ij1}^{\dagger} = b_{i1\uparrow}^{\dagger} 
b_{j2\uparrow}^{\dagger} \; \; \; \; \; \; \; \; \; \; \; \; \; \;  \; \; \; \; \; \; \;
\; \; \; \; \; for \; \; \; m=1, \\ 
B_{ij0}^{\dagger} = \frac{1}{\sqrt{2}} \left( 
b_{i1\uparrow}^{\dagger} b_{j2\downarrow}^{\dagger}  + 
b_{i1\downarrow}^{\dagger} b_{j2\uparrow}^{\dagger} \right) \; \; \; 
for \; \; \; m=0, \\  
B_{ij-1}^{\dagger} = b_{i1\downarrow}^{\dagger}
b_{j2\downarrow}^{\dagger} \; \; \; \; \; \; \; \; \; \; \;  \; \; \; \; \; \;
\; \; \; \; \; \; for \; \; \; m=-1. 
\end{array} 
\right. 
\label{24}
\ee 
In other words, the local triplet-pair creation operators are composed of
projected creation operators, one taken for site $i$, the other for the
neighboring site $j$. 
Thus, the Hamiltonian (\ref{22}) has a form similar to that for $t-J$ model
\cite{11} except we have here the triplet and {\it interband} pairing. 
This effective pairing Hamiltonian should be compared with the standard form of
the two-band Hamiltonian expressed through the spin and pseudospin operators,
respectively 
\be 
{\bf S}_i = \frac{1}{2} \sum_{l \s \s '} b_{il\s}^{\dagger} {\vec{\tau}}_{\s \s
'} b_{il \s '} , 
\label{25} 
\ee 
\be 
{\bf T}_i = \frac{1}{2} \sum_{\s l l'} b_{il\s}^{\dagger} {\vec{\tau}}_{l l'} 
 b_{il'\s } , 
\label{26} 
\ee 
where $\tau_{\xi \xi '}^{\alpha}$ are the elements of the Pauli matrix
$\tau^{\alpha}$, $\alpha = 1,2,3$. 
The effective Hamiltonian has then the form 
\be 
{\cal{H}} = \sum_{ijl\s}{'} t_{ij} b_{il\s}^{\dagger}b_{jl\s} - \frac{2}{K-J}
\sum_{<ij>} t_{ij}^2 \left( {\bf S}_i \cdot {\bf S}_j + \frac{3}{4} n_i n_j
\right) \left( \frac{1}{4} n_i n_j - {\bf T}_i \cdot {\bf T}_j \right) + ..., 
\label{27} 
\ee 
where the antiferromagnetic kinetic exchange part has not been included. 
This form is equivalent to the form (\ref{22}). 
In effect, the multiband model in the strong-correlation limit may exhibit
spin-triplet pairing (with $<B_{ijm}^{\dagger}> \neq 0$), spin ordering (with
$<S_{il}^z> \neq 0$), and the orbital ordering (with $<T_l^z> \neq 0$), or two
of the orderings simultaneously. 

One can solve the {\it ferromagnetic t-J model} (\ref{22}). 
This has been performed \cite{12} within the slave-boson formalism \cite{13} 
in which 
the projected fermion operators are decomposed into pseudofermion operator $f$
and the boson operator $b$, namely $b_{il\s}^{\dagger} = f_{il\s}^{\dagger}
b_i$. 
In effect, the effective Hamiltonian takes the form 
$$
{\cal{H}} = \sum_{ijl\s}{'} t_{ij}f_{il\s}^{\dagger}f_{jl\s} b_j^{\dagger}b_i -
\frac{2t^2}{K-J} \sum_{<ij><jk} F_{ijm}^{\dagger} F_{jkm} $$
\be 
+ \sum_{i} \lambda_i \left( b_i^{\dagger}b_i + \sum_{l\s}
f_{il\s}^{\dagger}f_{il\s} - 1 \right) - \mu \left( \sum_{il\s} 
f_{il\s}^{\dagger}f_{il\s} - N_e \right) , 
\label{28} 
\ee 
where 
$\lambda_i$ is the constant expressing the constraint 
\be 
b_i^{\dagger}b_i + \sum_{l\s} f_{il\s}^{\dagger}f_{il\s} = 1, 
\label{29} 
\ee 
and $F_{ijm}^{\dagger}$ operators have the same form as $B_{ijm}^{\dagger}$
with $b_{il\s}^{\dagger}$ being replaced by $f_{il\s}^{\dagger}$. 

When constructing the phase diagram for the system described by the effective
Hamiltonian (\ref{25}) we have to consider also a possibility of appearance of
ferromagnetism coexistent with the alternant (antiferromagnetic) ordering
(AFO). 
In Fig.5 we have plotted a simpler version of the phase diagram, on which we
have marked only saturated ferromagnetic phase (SC),  superconducting phase
labelled as S, paramagnetic metallic PM and paramagnetic superconducting (PM
and PS) phases, respectively, as well as the coexisting phases: AFO-SF and
S-SF. 
This phase diagram is for quarter filled, doubly degenerate band $(n=1, d=2)$, 
 with a constant density of states. 
The superconductivity is stable for low values of $U$ and a rather strong Hund's
rule coupling $J$. 
All the lines determining phase border lines mark the first-order phase
transition lines. 
The AFO-SF state is a Mott insulating state, whereas the remaining phases are
metallic. 
Hence the transition from AFO-SF to S-SF phase is also an insulator to metal
transition. 
The system is ferromagnetic on both sides of the transition and this transition
is possible only in the orbitally degenerate systems with non-half-filled band
configuration. 
This type of transition complements the standard canonical Mott transition, 
which take 
place from the antiferromagnetic insulator to either antiferromagnetic or
paramagnetic metal.

In order to see the regimes of  stability of the AFO and the SF states (and
their coexistence) with respect to the spin-triplet superconducting state as a
function of the band filling we have plotted in Fig.6 the phase diagram
involving the magnetic and orbitally ordered phases.  
The dashed lines represent 
a continuous phase transformation. 
The antiferromagnetic orbital ordering is stable only within $15\%$ filling
difference from the quarter filling. 
The nature of the transition evolves with 
the increasing ratio $J/U$. 
The slave boson approach presented in Refs. \cite{5} and \cite{7} has been used
to obtain the results valid for arbitrary $U$ and $J$. 
One should mention that the metallic AFO state is not in conflict with the
spin-triplet superconducting state. 
This question requires a separate discussion.

In Fig.7 we have shown (the upper panel) the regimes of the existence of 
various 
superconducting states (explained below) and the temperature $T_{RVB}$ below
which the gap parameters $<f_{i1\s}^{\dagger}f_{j2\s '}>$ are nonzero, as well
the temperature $T_D$ below which the slave bosons condense. 
Since the physical superconducting gap is $\sim <B_{ijm}^{\dagger}> \sim 
<F_{ijm}^{\dagger}><b_ib_j>$, the nonzero critical temperature for the
superconductivity is realized for $0.1 \lesssim n < 1$. 
Note that in this Figure 7 the exchange integral is defined as $J \equiv
2t^2/(K-J_H)$, where $J_H$ is the Hund's rule coupling (taken as $J$ in all
preceding discussion). 
So, a pure suerconducting phase should appear in the regime, in which
both AFO and SF states are absent and $T_{RVB} \neq 0$, i.e. for 
$0.60 \lesssim n \lesssim 0.85$. 

The superconducting phases specified in the upper panel of Fig.7 are defined
through the wave-vector $({\bf q})$ dependent two-dimensional representation 
of the superconducting gap ${\hat{\Delta}}_{{\bf q}}$ as follows 
\be 
{\hat{\Delta}}_{{\bf q}} = 2 {\hat{\Delta}} \left( \cos q_x + e^{i\Theta} \cos
q_y \right) 
\label{30} 
\ee 
with
\be 
{\hat{\Delta}} \equiv i \left( {\bf d} \cdot \tau \right) \s_y = 
\left( 
\begin{array}{cc}
-d_x+id_y, & d_z \\ 
d_z, & d_x +id_y 
\end{array} 
\right) 
\label{31} 
\ee 
being the standard form of the spin-triplet gap \cite{14}.  
For $\Theta = 0$ we have extended s-wave pairing, whereas for $\Theta = \pi$ we
have a d-wave pairing. 
We have selected a two-dimensional case for the numerical analysis since
$Sr_2RuO_4$ is regarded as a quasi-two-dimensional system. 
The qualitative features of the phase diagram do not depend much on the $J/t$
ratio if only $J$ is substantially smaller than $t \equiv |t_{<ij>}|$. 

Summarizing, in this Section we have discussed a coexistence of the saturated
ferromagnetic and A1 superconducting state for the quarter filled band in the
strongly correlated regime, as well as its competition with the AFO-SF state. 
The transformation for $n=1$ of the AFO-SF insulating state into S-SF metallic
state for $J \geq U/3$ is accompanied by a closure of the Mott-Hubbard
\cite{7}.  
Hence, not only the appearance of the spin-triplet superconductivity, but also
the metallicity are both induced by the Hund's rule coupling $J$ in this case. 
As we have obtained a stable spin-triplet superconductivity (with an isotropic
gap ($\Delta_{{\bf q}} = \Delta$) in the weak-coupling (Hartree-Fock BCS
limit), as well as in the strong-correlation limit (this time with {\bf
q}-dependent gap), we can say that this type of superconductivity is a generic
phenomenon of our model of correlated electrons with orbital degeneracy, at
least in the mean-field approximation. 
It would be desirable to include to discuss the stability of the present
results against the quantum Gaussian fluctuations.

\vskip1.0cm 

\noindent {\Large {\bf 7. Concluding remarks
}} 

\vskip0.5cm 

In this article we have reviewed briefly the mean-field approach to the
spin-triplet superconductivity in orbitally degenerate narrow band systems that
is induced by the local ferromagnetic (Hund's rule) exchange. 
Both weak (Hartree-Fock)- and strong-correlation regimes were discussed and
concrete quantitative results have been presented for the case of a doubly
degenerate band. 
The hybridization of bands has been neglected and therefore the microscopic
parameters ($W$, $\e_F$, $U$, and $J$) represent effective values. 
Nonetheless, explicit calculations for a hybridized model should be performed
and this should allow for an analysis of concrete materials. 
Hybridization will introduce a {\bf q}-dependence to the superconducting gap
even in the weak-coupling regime. 

The Hund's rule coupling allows for treating a weak itinerant-electron
ferromagnetism and real-space spin-triplet pairing on equal footing within a
single mechanism. 
However, the spin fluctuation contribution should be included to see their
relative role in scattering the carriers (particularly near $T_C$) and
providing the pairing in the temperature regime $T \lesssim T_S \ll T_C$.

\vskip1.0cm 

\noindent {\Large {\bf Acknowledgement
}} 

\vskip0.5cm 

The authors are grateful to Leszek Spa\l{}ek for his technical help and a
constant encouragement. 
The discussions with Karol Wysoki\'nski, Mark Jarrell, Ben Powell, Canio Noce,
and Mario Cuoco are appreciated. 
The work was supported by the State Committee for Scientific Research (KBN) of
Poland.

\newpage

{\large {\bf Figure Captions}}

\vskip0.5cm

\noindent {\bf Fig.1.} Phase diagram involving the spin-triplet superconducting A and A1
states in an applied magnetic field $B$ for a two-band model at quarter filling
and with a constant density of states. 
The dot marks the transition from anisotropic A phase (with $\Delta_1 \equiv
\Delta_{\ua \ua} > \Delta_{-1} \equiv \Delta_{\da \da}$ to A1 phase
$(\Delta_{\da \da} =0)$. 

\vskip0.5cm

\noindent {\bf Fig.2.} Upper panel: Magnetic moment $<S_l^z> \equiv {\bar{S^z}}$ in the field
for the same situation as in Fig.1; lower panel: Field dependence of the
superconducting gaps as marked. Inset: Field dependence of the chemical
potential in the A and A1 phases.

\vskip0.5cm

\noindent  {\bf Fig.3.} Magnetic moment $<S_l^z>$ induced by the spin-triplet pairing below the
Stoner threshold (marked as Stoner criterion, $J/W \approx 0.125$). 
The effect is spectacular when the system approaches the critical point. 
Inset: magnetic moment vs. $J/W$ if the spin-triplet pairing were absent (the
para(PM)- to ferro(FM)- magnetic transition is discontinuous at the Stoner
point for the constant density of states selected to illustrate the {\it
hiding} of the Stoner critical point in the paired state.

\vskip0.5cm

\noindent  {\bf Fig.4.} Critical magnetic field for $A \rightarrow A1$ transition (solid line)
and the field saturating the moment (i.e. making ${\bar{S^z}} = 1/4$). 
The Stoner threshold is also marked. 
Note that the threshold for an apprearance of the polarized paired (A1) state
does not coincide with the onset of saturated ferromagnetic (SF) state. 
The A phase disappears at the Stoner threshold. 

\vskip0.5cm

\noindent  {\bf Fig.5.} Magnetic phase diagram on the plane $U-J$ for a quarter filled doubly
degenerate band. 
The symbols label the corresponding phases: PM - paramagnetic metallic, 
AFO-SF - antiferromagnetic orbital ordering of saturated ferromagnet, PS -
paramagnetic spin-triplet superconductors, S-SF - A1 superconducting phase
coexisting with saturated ferromagnetism. 
All the lines mark discontinuous phase transitions at temperature $T=0$. 

\vskip0.5cm

\newpage

\noindent  {\bf Fig.6.} The evolution of the AFO-SF and SF states with the increasing Hund's
rule coupling $J$ (the phase labelling is the same as in Fig.5), plotted as a
function of band filling \\ $n=\sum_{l\s} <n_{il\s}>$, for a doubly-degenerate
band with a constant density of states. 

\vskip0.5cm 

\noindent  {\bf Fig.7.}  Upper panel: Phase diagram for $n \rightarrow 1$ for a
quasi-two-dimensional model involving spin-triplet pairing with {\bf
q}-dependent gap obtained in the strong-correlation regime. 
$T_D$ represents the temperature of the Bose condensation of auxiliary bosons,
$T_{RVB}$ the formation of a BCS-like state (the RVB state) for 
pseudofermions. 
Lower panel: The condensation temperatures in a wide range of the band filling.

\newpage
\begin{figure}
\epsfxsize15cm
\epsffile{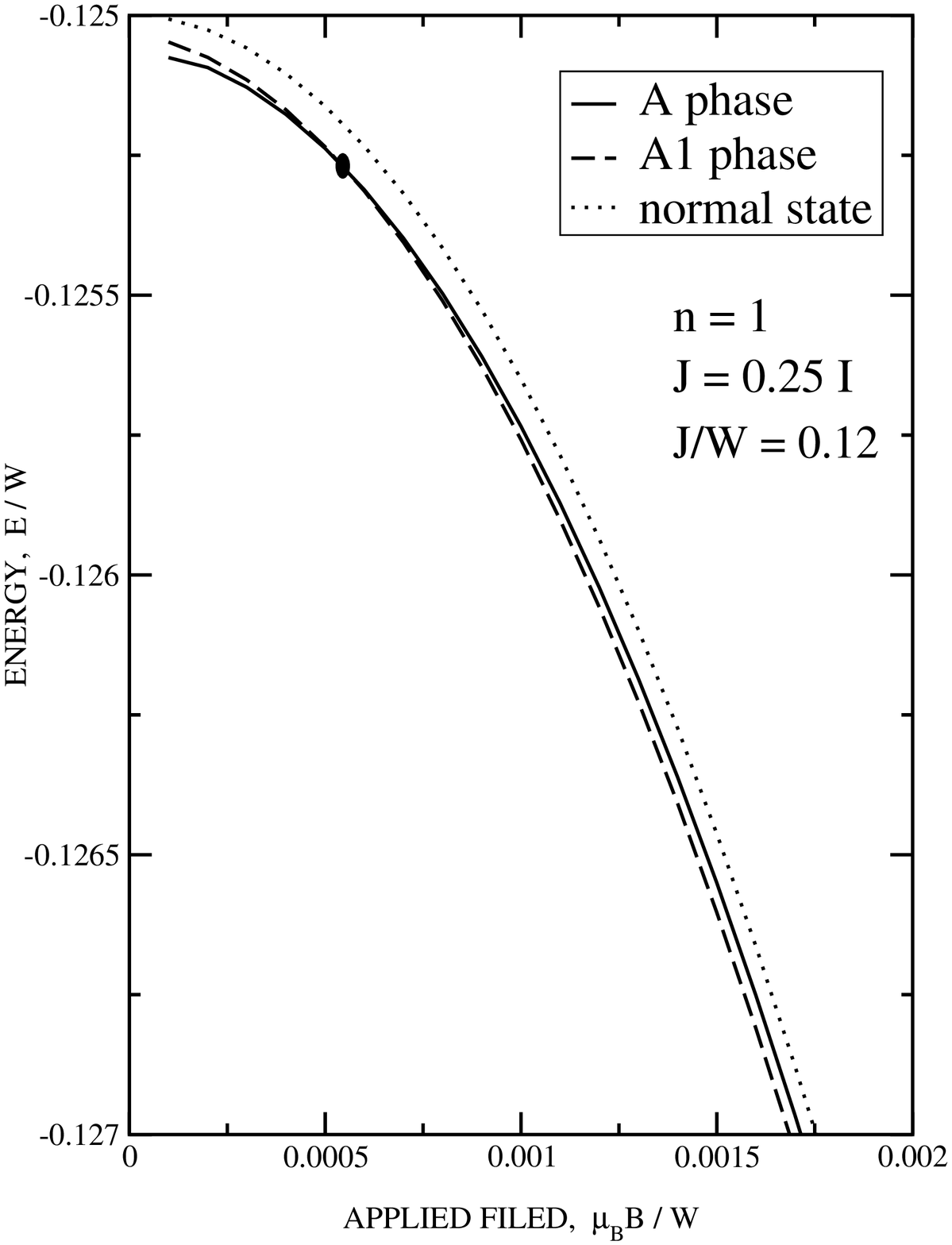}
\caption{}
\end{figure}

\newpage
\begin{figure}
\epsfxsize15cm
\epsffile{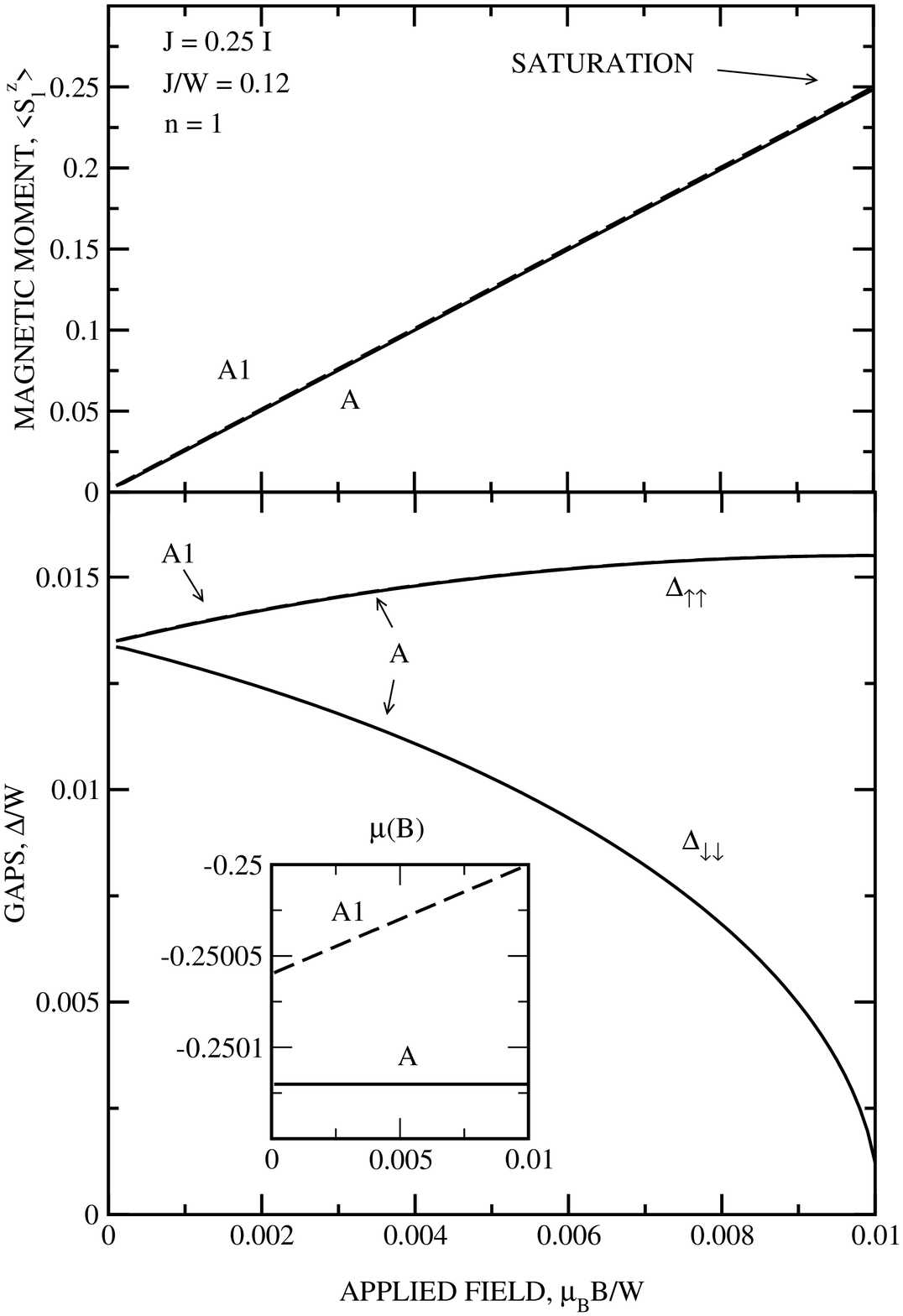}
\caption{}
\end{figure}

\newpage
\begin{figure}
\epsfxsize15cm
\epsffile{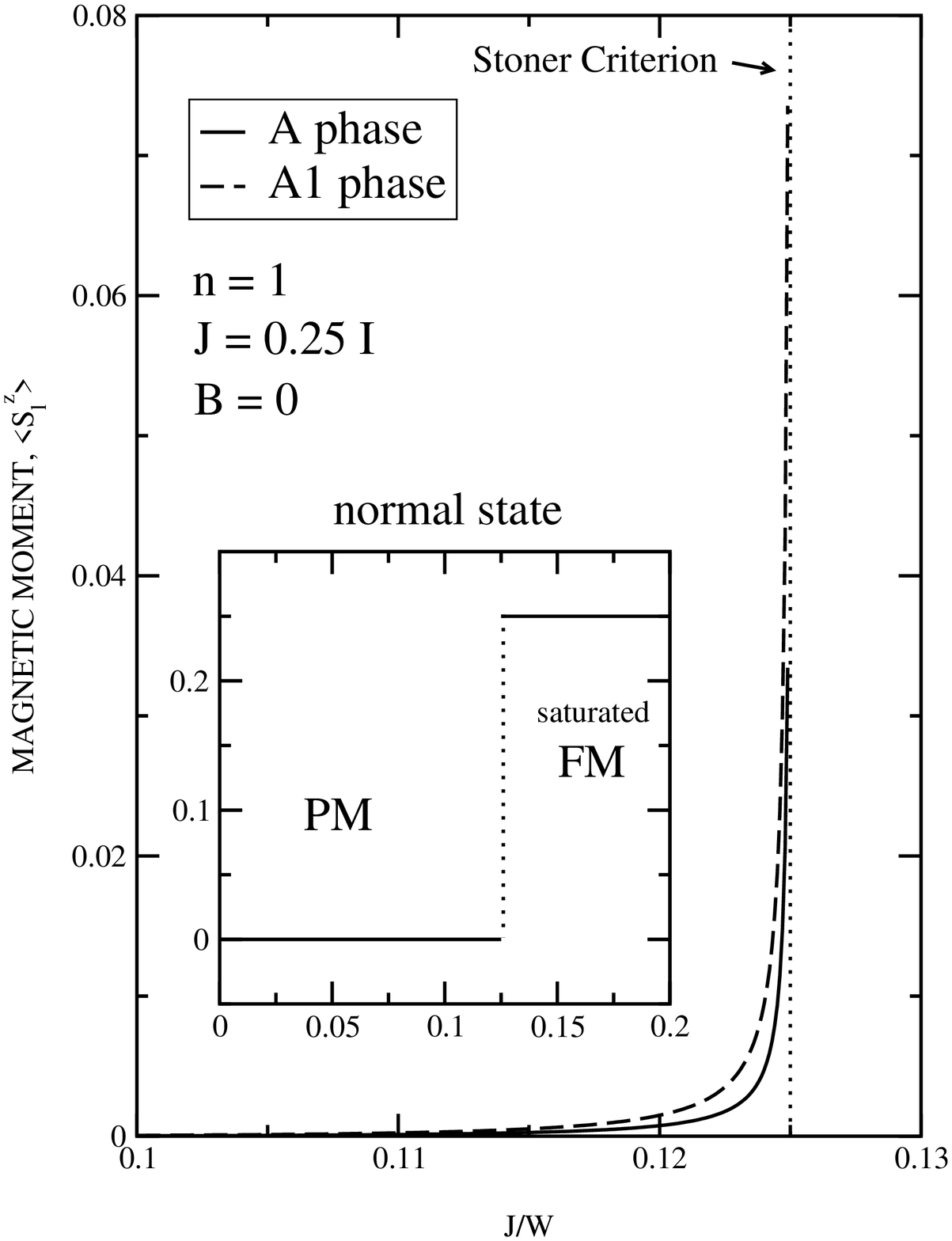}
\caption{}
\end{figure}

\newpage
\begin{figure}
\epsfxsize15cm
\epsffile{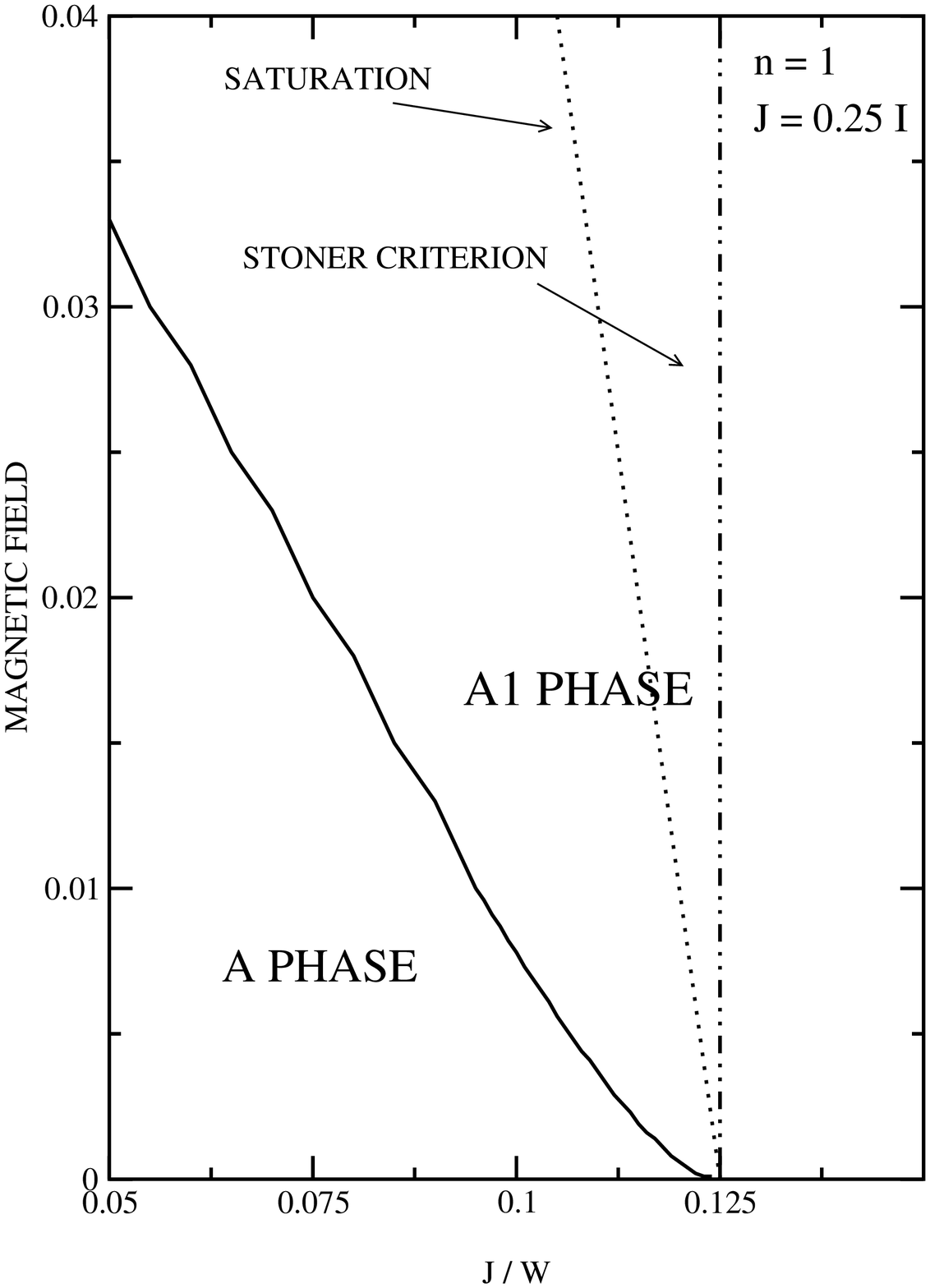}
\caption{}
\end{figure}

\newpage
\begin{figure}
\epsfxsize15cm
\epsffile{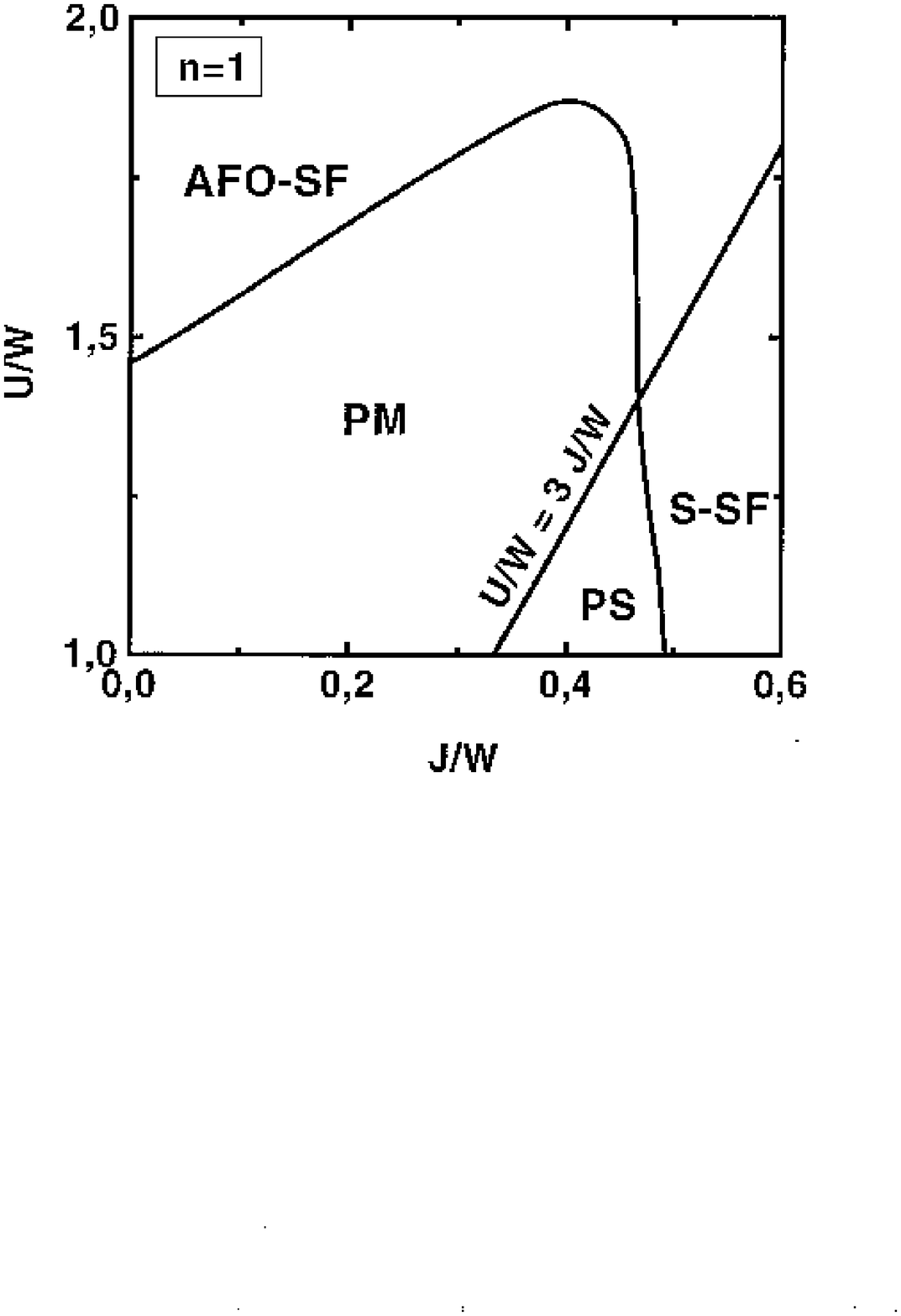}
\caption{}
\end{figure}

\newpage
\begin{figure}
\epsfxsize15cm
\epsffile{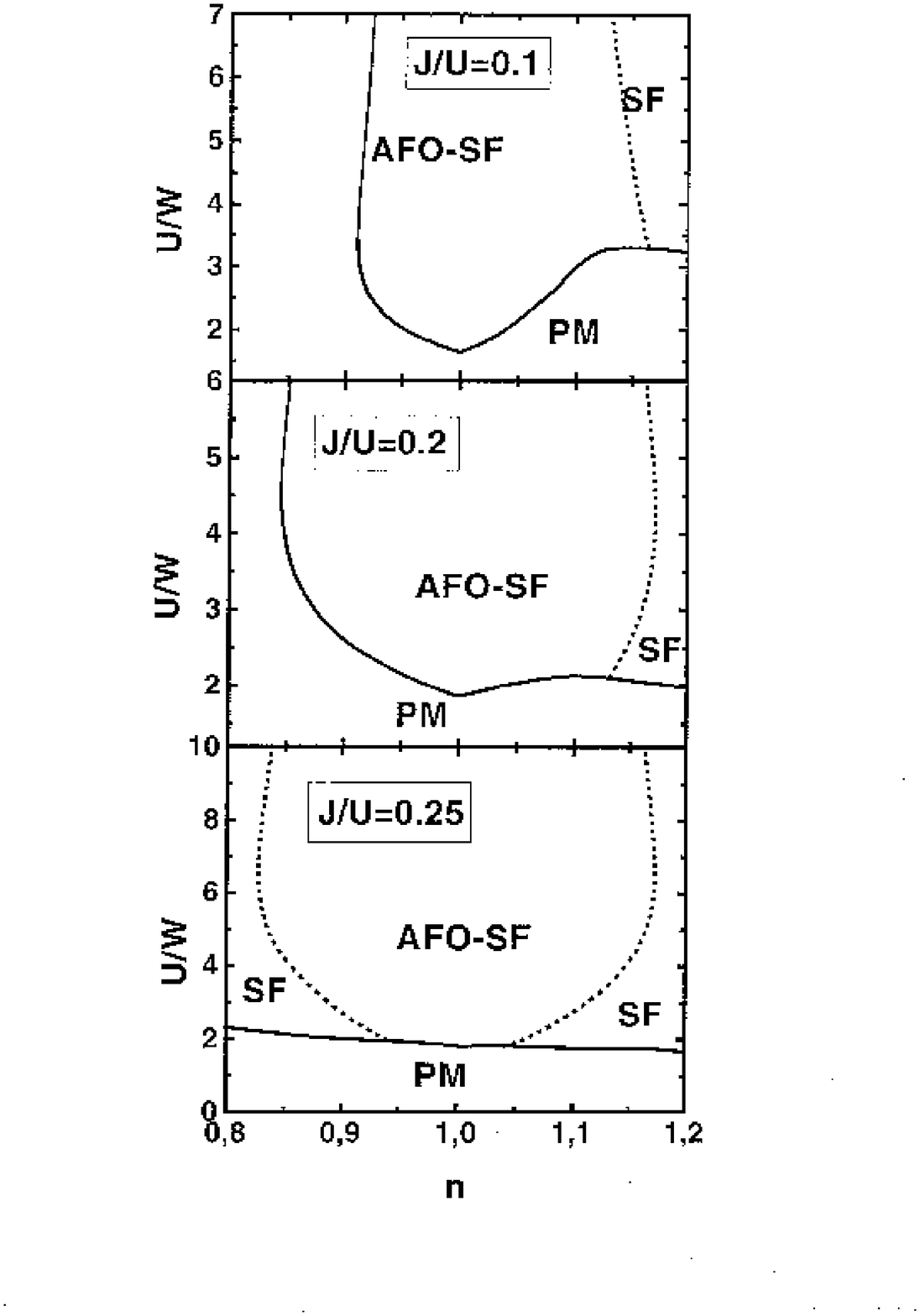}
\caption{}
\end{figure}

\newpage
\begin{figure}
\epsfxsize15cm
\epsffile{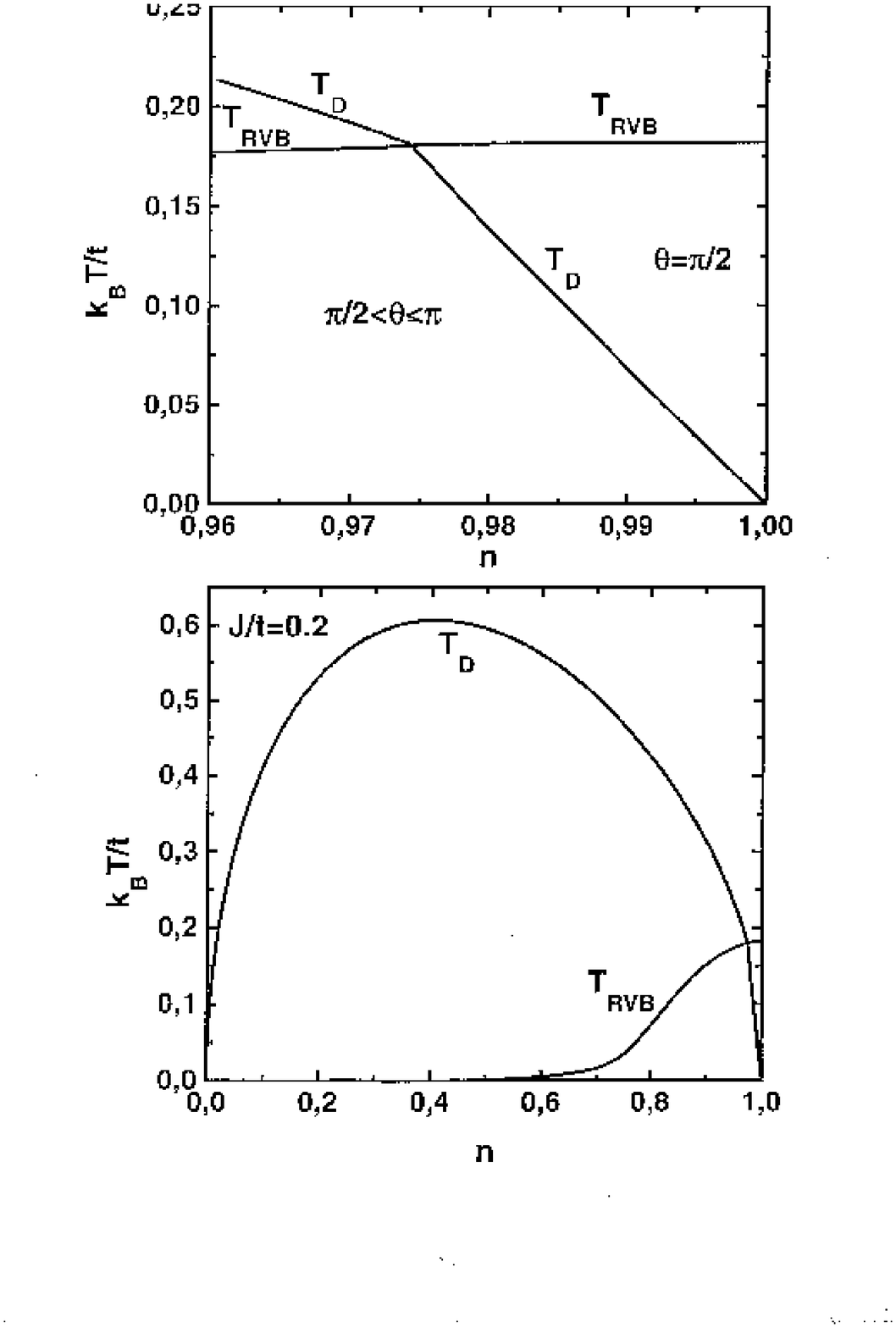}
\caption{}
\end{figure}

\end{document}